\documentclass[preprint]{vgtc}               

\ifpdf
  \pdfoutput=1\relax                   
  \usepackage{graphicx}                
  \DeclareGraphicsExtensions{.pdf,.png,.jpg,.jpeg} 
\else
  \usepackage{graphicx}                
  \DeclareGraphicsExtensions{.eps}     
\fi%

\graphicspath{{figures/}{pictures/}{images/}{./}} 
\usepackage{xcolor}
\usepackage{amsmath}
\usepackage{bbm}
\usepackage{subcaption}
\usepackage{bm}
\usepackage{microtype}                 
\PassOptionsToPackage{warn}{textcomp}  
\usepackage{textcomp}                  
\usepackage{mathptmx}                  
\usepackage{times}                     
\usepackage{cite}                      
\usepackage{tabu}                      
\usepackage{booktabs}                  

\newcommand{\fix}[1]{\textcolor{black}{#1}}
\usepackage[normalem]{ulem}
\onlineid{1129}

\vgtccategory{Research}

\preprinttext{To appear in the IEEE VIS 2021 conference.}

\title{Uncertainty Visualization of the Marching Squares and \\Marching Cubes Topology Cases}

\author{Tushar M. Athawale\thanks{e-mail: tushar.athawale@gmail.com} %
\and Sudhanshu Sane\thanks{e-mail: ssane@sci.utah.edu} %
\and Chris R. Johnson\thanks{e-mail: crj@sci.utah.edu}}
\affiliation{\scriptsize Scientific Computing \& Imaging (SCI) Institute, University of Utah}

\abstract{Marching squares (MS) and marching cubes (MC) are widely used algorithms for level-set visualization of scientific data. In this paper, we address the challenge of uncertainty visualization of the topology cases of the MS and MC algorithms for uncertain scalar field data sampled on a uniform grid. The visualization of the MS and MC topology cases for uncertain data is challenging due to their exponential nature and the possibility of multiple topology cases per cell of a grid. We propose the {\em topology case count} and {\em entropy-based techniques} for quantifying uncertainty in the topology cases of the MS and MC algorithms when noise in data is modeled with probability distributions. We demonstrate the applicability of our techniques for independent and correlated uncertainty assumptions. We visualize the quantified topological uncertainty via color mapping proportional to uncertainty, as well as with interactive probability queries in the MS case and entropy isosurfaces in the MC case. We demonstrate the utility of our uncertainty quantification framework in identifying the isovalues exhibiting relatively high topological uncertainty. We illustrate the effectiveness of our techniques via results on synthetic, simulation, and hixel datasets.%
} 

\CCScatlist{
  \CCScatTwelve{Human-centered computing}{Visu\-al\-iza\-tion}{Visualization application domains}{Scientific visualization}
}

\begin{document}


\firstsection{Introduction}

\maketitle
Level-sets are a fundamental surface-based visualization technique used to help understand complex scientific datasets. The marching squares (MS) and marching cubes (MC) algorithms~\cite{Lorensen:1987:MCA} are extensively used by the scientific visualization community for level-set extraction from univariate scalar fields in two and three dimensions, respectively. Both algorithms reconstruct the topology and geometry of level-sets by stepping through the cells of a Cartesian grid on which data are sampled. We study the uncertainty in the topology reconstruction step of the MS and MC algorithms when the data sampled at grid vertices are not fixed or are uncertain. 


Analyzing the effects of data uncertainty on visualization algorithms has proved effective in applications such as medical~\cite{RISTOVSKI201460} and geo-spatial data analysis~\cite{Pang2001VisualizingUI}, helping to identify and reduce visual misrepresentations of underlying data. Thus, uncertainty visualization has been recognized as a top research challenge~\cite{JohnsonSanderson2003, zuk:2007:VUR, PotterRosenJohnson2012, BrodlieAllendesLopes2012, BonneauHegeJohnson2014, TA:KamalDhakal2021}. In the context of level-sets, multiple uncertainty visualizations, such as spaghetti plots~\cite{potter2009}, contour boxplot~\cite{WhitakerMirzargarKirby2013}, probabilistic marching cubes~\cite{PothkowWeberHege2011, PothkowHege2013}, a closed-form level-set uncertainty framework~\cite{AthawaleEntezari2013, AthawaleSakhaeeEntezari2016, AthawaleJohnson2019}, and confidence level-sets~\cite{zehner2010visualization,sane2021fcls} have previously been proposed. In this work, we propose a new framework for uncertainty visualization of level-sets to study the uncertainty arising in the MS and MC topology cases due to noisy scalar field data.

A few recent studies have analyzed the effects of uncertainty in data on level-sets at the grid-cell granularity for the MS and MC algorithms, but did not study their topology variations. Specifically, P\"{o}thkow et al.~\cite{PothkowWeberHege2011} computed the \fix{probability of level-set passing through a cell, i.e., a {\em level-crossing probability},} for each cell of a Cartesian grid for uncertain data, but no analysis was presented to quantify the level-set topology variations within a cell. Similarly, Athawale et al.~\cite{AthawaleSakhaeeEntezari2016} studied the most probable topology per cell for uncertain data and the uncertainty in inverse linear interpolation (Ilerp uncertainty) along cell edges, but did not quantify the uncertainty in topology. We fill the gap in the prior work by proposing a novel framework to quantify and visualize the uncertainty in the topology reconstruction step of the MS and MC algorithms.    

\begin{figure}
\begin{center}
  \includegraphics[width=\linewidth]{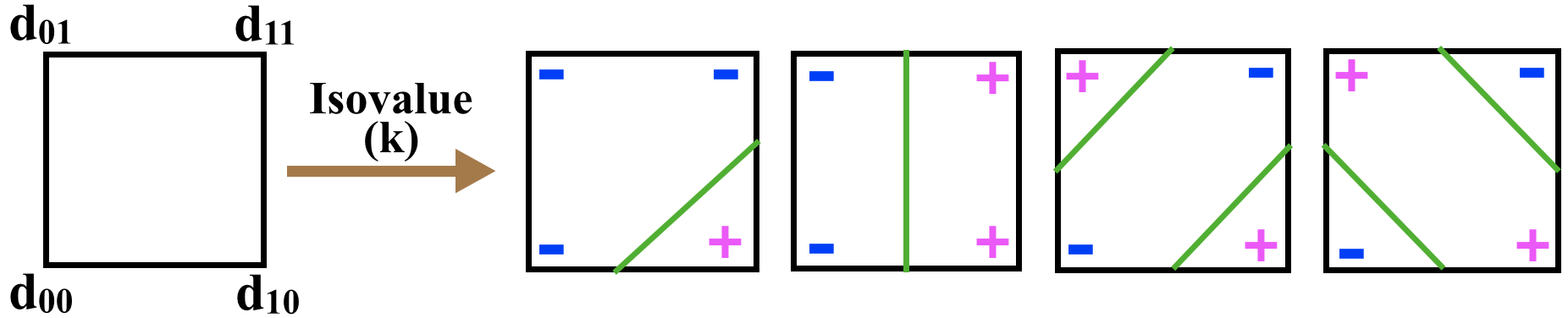}
  \vspace{-5mm}
\end{center}
\vspace{-3mm}
\caption{Depiction of the MS topology cases: If $\text{data }(d_{ij}) \ge \text{isovalue }(k)$, the vertex is marked $+$; otherwise, it is marked $-$. The level-set topology (green) is shown for different sign configurations.}
\label{fig:msTopology}
\vspace{-5mm}
\end{figure}

\fix{Fig.~\ref{fig:msTopology} illustrates the MS topology cases~\cite{Lorensen:1987:MCA} for a single 2D cell.}  As depicted in Fig.~\ref{fig:msTopology}, let $d_{xy}$ denote fixed data sampled at the cell vertices of a 2D Cartesian grid, where $x$ and $y$ denote the local cell coordinates. Depending upon the isovalue $k$, a cell vertex is assigned a positive sign if $d_{xy} \ge k$; otherwise, it is assigned a negative sign. The isocontour topology (green line segments in Fig.~\ref{fig:msTopology}) is then extracted within a cell such that it isolates the positive from the negative vertices~\cite{Lorensen:1987:MCA}. Since there are four vertices and each vertex can attain a positive or negative sign, there are $2^4$, i.e., $16$, possible MS topology cases per cell. The $16$ MS topology cases may be summarized using the four cases, as shown in Fig.~\ref{fig:msTopology}, by considering the symmetry of level-set orientations. The MC algorithm has $2^8$, i.e., $256$, topology cases since each cell has eight vertices. 



When data at the 2D cell vertices are uncertain, each grid cell can attain any number of the MS topology cases between $1$ and $16$. This behavior is different from the certain data scenario, where each cell attains only one of the $16$ topology cases. In our proposed framework, we build upon the frameworks proposed by P\"{o}thkow et al.~\cite{PothkowWeberHege2011} and Athawale et al.~\cite{AthawaleSakhaeeEntezari2016}, and utilize an entropy-based quantification, similar to~\cite{Athawale:2020:AJPW}, for uncertainty visualization of the level-set topology cases. The entropy of probability distributions characterizing gradient flow uncertainty was utilized by Athawale et al.~\cite{Athawale:2020:AJPW} to quantify the uncertainty in Morse complex segmentation boundaries. In our work, we quantify the uncertainty of the MS and MC topology cases by computing the entropy of their probability distributions.

\paragraph{Contributions:}
In this work, we present the topology case count and entropy-based statistical frameworks (Sec.~\ref{sec:methods}) for uncertainty quantification and visualization of the MS and MC topology cases. We demonstrate the applicability of our frameworks for independent and correlated random field assumptions. Our results (Sec.~\ref{sec:results}) demonstrate that our proposed techniques more effectively capture local topological uncertainty of level-sets compared to spaghetti plots~\cite{potter2009}, Ilerp uncertainty~\cite{AthawaleSakhaeeEntezari2016}, and probabilistic marching squares/cubes~\cite{PothkowWeberHege2011}. We show how our statistical framework can be leveraged for identifying the isovalues with relatively high or low topological uncertainty.

\section{Background}
\label{sec:background} 

We briefly discuss the vertex-based classification~\cite{AthawaleSakhaeeEntezari2016} and probabilistic marching cubes~\cite{PothkowWeberHege2011} frameworks, which are the building blocks of our proposed uncertainty quantification framework.


\subsection{Vertex-Based Classification}
\label{sec:vbc}
The vertex-based classification predicts the most probable MS or MC topology case for each cell of a Cartesian grid, where uncertain data are assumed to be sampled from independent probability distributions. In Fig.~\ref{fig:msTopology}, we denoted the fixed data at cell vertices with a variable $d_{xy}$. When data have uncertainty, we denote the data at cell vertices by a random variable $D_{xy}$.  Let $\text{\fix{pdf}}_{D_{xy}}$ be the probability distribution at each vertex estimated from sample data. For the isovalue $k$, let $D^+_{xy} = Pr(D_{xy} \ge k)$, i.e., the probability of a cell vertex (x,y) attaining a positive vertex sign. Similarly, let $D^-_{xy} = Pr(D_{xy} < k)$. In a vertex-based classification, the vertex is predicted as positive if $D^+_{xy} \ge D^-_{xy}$; otherwise, it is predicted as negative. The predicted signs, therefore, indicate the single most probable MS or MC topology case for each cell of an uncertain scalar field. 



\subsection{Probabilistic Marching Cubes}
\label{sec:pmc}
The probabilistic marching cubes estimates the level-crossing probability for each cell of a Cartesian grid, where uncertain data are assumed to be sampled from multivariate Gaussians. For a 2D version of the probabilistic marching cubes, i.e., the probabilistic marching squares, let $D= (D_{00}, D_{01}, D_{10}, D_{11})^T$ denote a random variable representing uncertain data at the 2D cell vertices. The random variable is assumed \fix{to have} a multivariate Gaussian distribution with the sample mean $\tilde{\bm{\mu}} = ({\tilde{\mu}_{D_{00}}, \tilde{\mu}_{D_{01}}, \tilde{\mu}_{D_{10}}, \tilde{\mu}_{D_{11}}})^T$ and sample covariance matrix $\tilde{\sum} = E[(D-\tilde{\bm{\mu}})(D-\tilde{\bm{\mu}})^T]$. The $N$ samples are then drawn from the distribution $\mathcal{N}(\tilde{\mu},\tilde{\sum})$. \fix{If the level-set with isovalue $k$ crosses a cell for $M$ number of samples, then the level-crossing probability for the cell is estimated as $\frac{M}{N}$}. The same approach is extendable to 3D. \fix{Note that the higher value of a sample count $N$ provides a more reliable estimation of the level-crossing probability.}



\section{Methods}
\label{sec:methods} 
\fix{We now describe our framework for the uncertainty quantification and visualization of the level-set topology cases.} For simplicity, we limit our descriptions to the 2D MS algorithm, but \fix{they are directly applicable} to the 3D MC algorithm. 

\subsection{Computing Topology Case Probability Distribution}
\label{sec:topoCaseDistribution}
\fix{In the uncertainty quantification step, we characterize the uncertainty of the MS topology cases by computing their probability distribution.} For the {\em independent random field assumption}, we leverage the vertex-based classification framework (Sec.~\ref{sec:vbc}) to compute the topology case probability distribution. First, we compute $D^+_{xy}$ and  $D^-_{xy}$ for each cell vertex. We then compute the probability for each of the $16$ MS topology cases per cell. For example, the probability of $D_{00}$, $D_{10}$, and $D_{11}$ being positive and $D_{01}$ being negative is equal to the product $D^+_{00} \cdot D^-_{01}\cdot D^+_{10} \cdot D^+_{11}$ because of the independence assumption. 

For the {\em correlated random field assumption}, we leverage the probabilistic marching cubes framework (Sec.~\ref{sec:pmc}) to compute the topology case probability distribution. First, we draw $N$ samples from a multivariate Gaussian distribution $\mathcal{N}(\tilde{\mu},\tilde{\sum})$. Next, we empirically compute the histogram of $16$ topology cases depending upon the topology case observed for each of the $N$ samples. \fix{Note that the Monte Carlo sampling for the correlated noise assumption results in approximate and expensive computations unlike the closed-form and fast computations for the independent noise models.}


%
 
\subsection{Topology Case Count Visualization}
\label{sec:topoCaseCount}
In the topology case count technique, we first compute the topology case probability distribution per grid cell, as described in Sec~\ref{sec:topoCaseDistribution}. \fix{We then derive the topology count field in which we count the number of topology cases per cell that have a probability greater than the user-specified lower probability threshold $t$, where $t \in [0,1]$. Setting a lower threshold provides users the flexibility to study uncertainty among topology cases with relatively high probability.} Let $C$ denote a discrete random variable representing the $16$ MS topology cases for a 2D cell $q$, and $\text{\fix{pdf}}_{C}(q)$ denote the topology case probability distribution for the cell $q$. \fix{Mathematically, the topology count for cell $q$ is equal to $\sum_{c=1}^{c=16} \mathbbm{1}{_{(Pr_{C = c}(q) > t)} (c)}$, where $\mathbbm{1}$ is the indicator function. Finally, we visualize the topology count field via colormapping. Note that for our experiments in Sec.~\ref{sec:results}, we set $t=0$.}

\subsection{Entropy-Based Uncertainty Visualization}
In the entropy-based technique, we first compute the topology case probability distribution per grid cell, as described in Sec~\ref{sec:topoCaseDistribution}. We then compute the Shannon entropy~\cite{Shannon1948} E(q) of the topology case probability distribution $\text{\fix{pdf}}_{C}(q)$ for each cell $q$ as: $E(q)=-\sum_{c=1}^{c=16} Pr_{C = c}(q) log_2 Pr_{C=c}(q)$. Finally, we visualize the entropy field via colormapping. The entropy of a topology case probability distribution quantifies the level of randomness of the topology cases within each cell, which may not be captured by the topology case count technique. Low entropy implies the relatively more deterministic nature of the topology cases, whereas high entropy implies the relatively more uncertain nature of the topology cases. Our entropy-based approach is inspired by the similar entropy-based approach proposed in~\cite{Athawale:2020:AJPW} for the visualization of uncertainty in Morse complexes. \fix{Our entropy-based framework can be used to identify the isovalues that exhibit relatively high or low topological uncertainty. Specifically, we visualize a boxplot of the entropy of grid cells with nonzero entropy/uncertainty for different isovalues to compare the topological uncertainty distribution of isovalues.}



\section{Results}
\label{sec:results} 

\begin{figure}[!ht]
\centering
\begin{subfigure}{0.45\linewidth}
\centering
\includegraphics[width=0.78\linewidth]{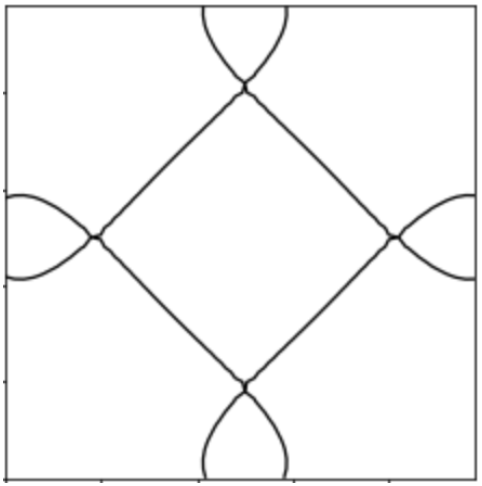}
\vspace{-2mm}
\caption{\fix{Level-set in an \\ image without noise}}
\label{fig:ackley_gt}
\end{subfigure}
\hspace{1mm}
\begin{subfigure}{0.45\linewidth}
\centering
\includegraphics[width=0.78\linewidth]{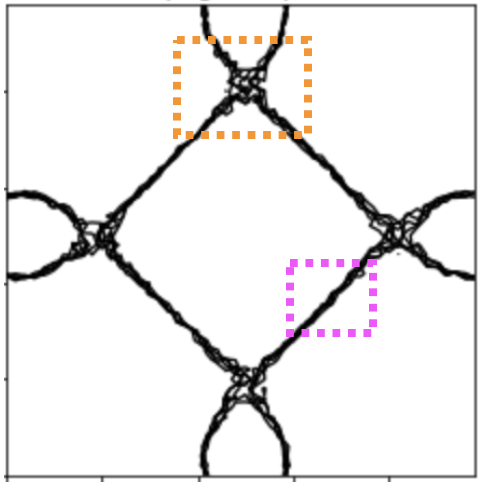}
\vspace{-2mm}
\caption{Spaghetti plot\\~\cite{potter2009}}
\label{fig:ackley_spaghetti}
\end{subfigure}

\vspace{-2mm}

\begin{subfigure}{0.49\linewidth}
\centering
\vspace{3.5mm}
\includegraphics[width=0.9\linewidth]{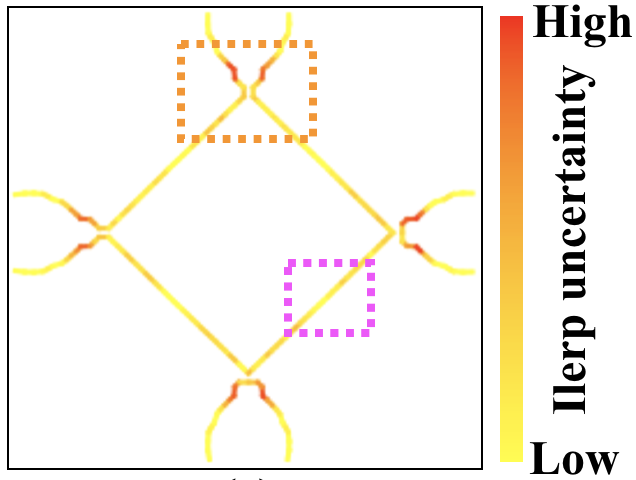}
\vspace{-2mm}
\hspace{-8mm}
\caption{Ilerp uncertainty \\visualization~\cite{AthawaleEntezari2013}}
\label{fig:ackley_ilerp}
\end{subfigure}
\begin{subfigure}{0.49\linewidth}
\centering
\vspace{3.5mm}
\includegraphics[width=0.95\linewidth]{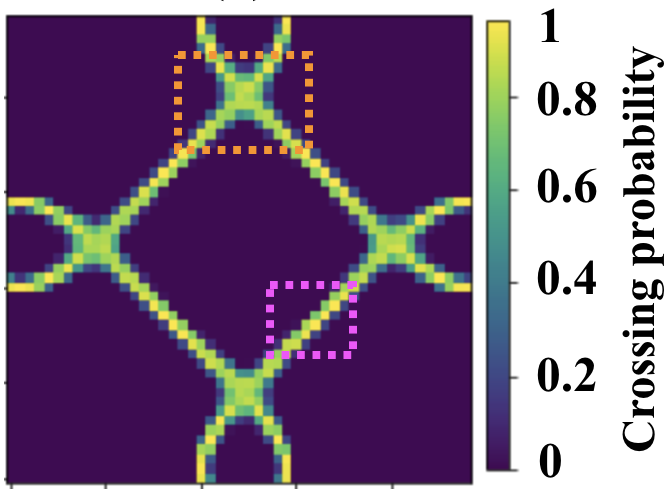}
\hspace{-8mm}
\vspace{-2mm}
\caption{Probabilistic marching \\squares~\cite{PothkowWeberHege2011}}
\label{fig:ackley_pms}
\end{subfigure}

\vspace{-2mm}

\begin{subfigure}{0.49\linewidth}
\centering
\vspace{3.5mm}
\includegraphics[width=0.9\linewidth]{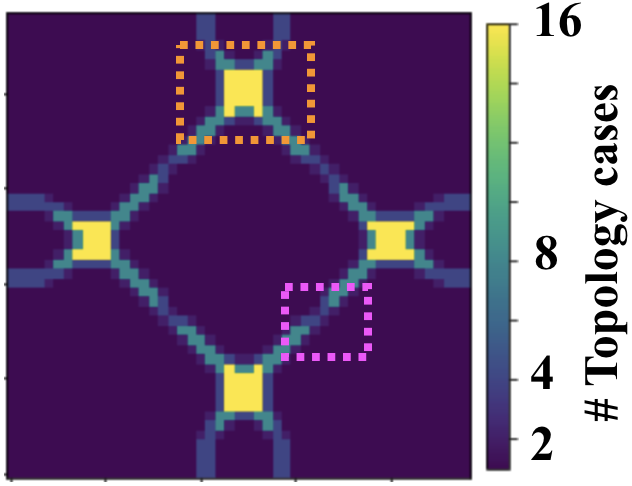}
\vspace{-2mm}
\hspace{-8mm}
\caption{Topology case count \\visualization}
\label{fig:ackley_topoCount}
\end{subfigure}
\begin{subfigure}{0.49\linewidth}
\centering
\vspace{3.5mm}
\includegraphics[width=0.95\linewidth]{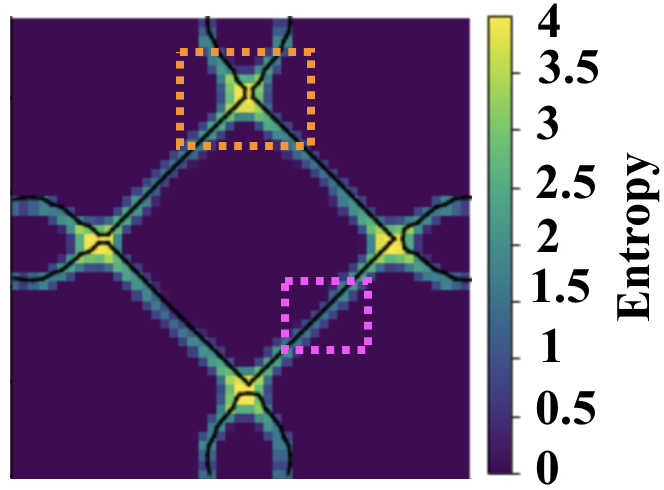}
\hspace{-8mm}
\vspace{-2mm}
\caption{Entropy-based\\ visualization}
\label{fig:ackley_entropy}
\end{subfigure}
\vspace{-2mm}
\caption{Level-set uncertainty visualizations for the Ackley dataset~\cite{Ackley1987} at isovalue $k=-3.25$ using the state-of-the-art techniques in images (b-d) and our proposed techniques in images (e-f). The orange and pink dotted boxes mark the positions with the relatively high and low topological variations, respectively.} 
\label{fig:ackleyVis}
\end{figure}

We demonstrate the effectiveness of our topology count and entropy-based level-set uncertainty visualizations in Fig.~\ref{fig:ackleyVis} via a synthetic experiment. The level-set for the \fix{Ackley dataset~\cite{Ackley1987}} is visualized in Fig.~\ref{fig:ackleyVis}a. We mix the dataset with \fix{uniform-distributed noise samples} to generate an ensemble.  Figs.~\ref{fig:ackleyVis}b-f visualize the results for the ensemble using multiple uncertainty visualization techniques for the independent uniform noise assumption, in which the mean and width of a distribution per vertex are estimated from the ensemble. 


\begin{figure}[!ht]
\begin{center}
  \hspace{-10mm}
  \includegraphics[width=0.8\linewidth]{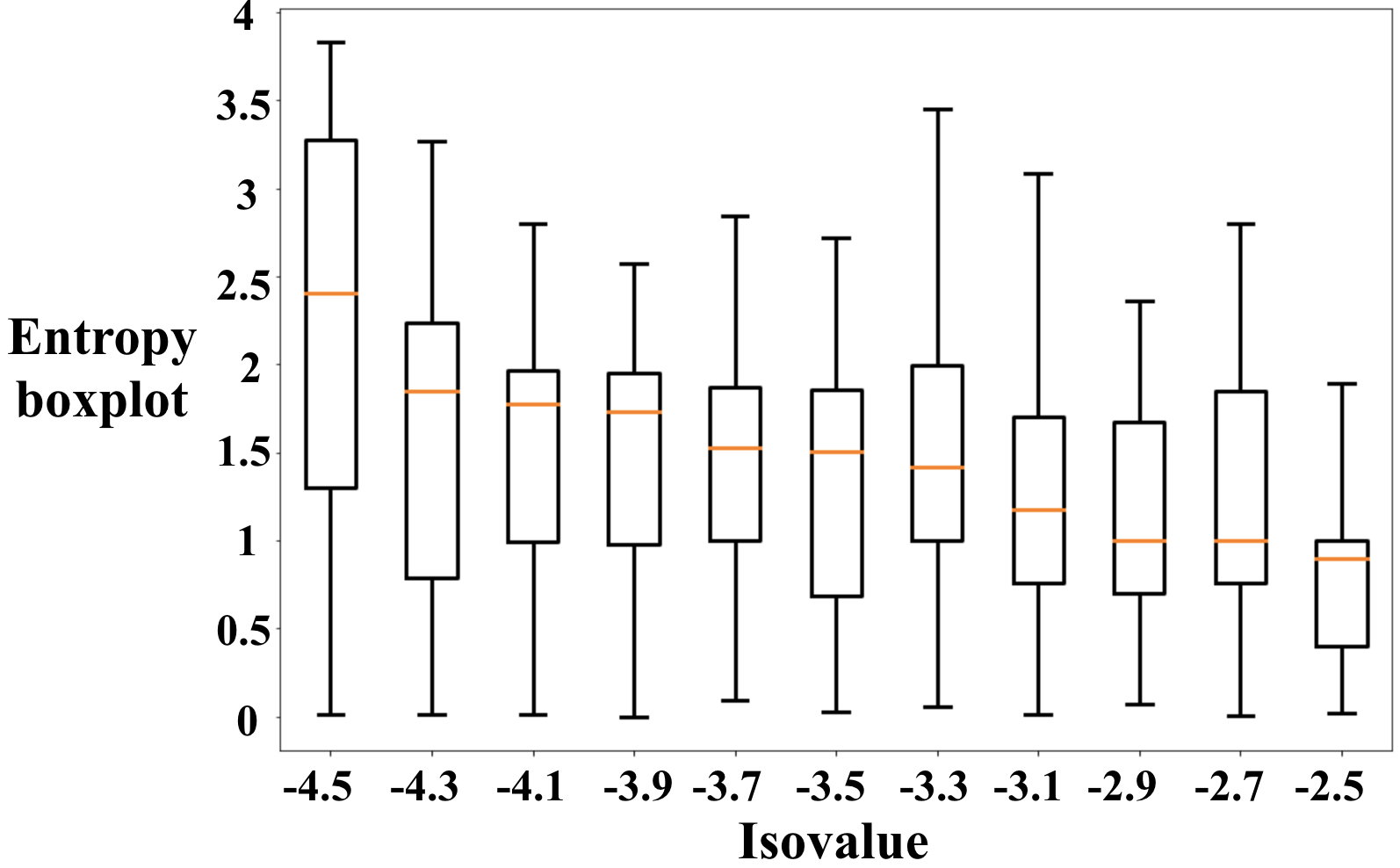}
\end{center}
\vspace{-6mm}
\caption{\fix{A boxplot of the entropy of grid cells with nonzero entropy is visualized per sampled isovalue to understand the topological uncertainty distribution of isovalues for the Ackley dataset.}}
\label{fig:entropyVsIsovalueAckleyVis}
\vspace{-5mm}
\end{figure}


\begin{figure*}[!ht]

\centering
\begin{subfigure}{0.18\linewidth}
\centering
\vspace{5.5mm}
\includegraphics[width=0.9\linewidth]{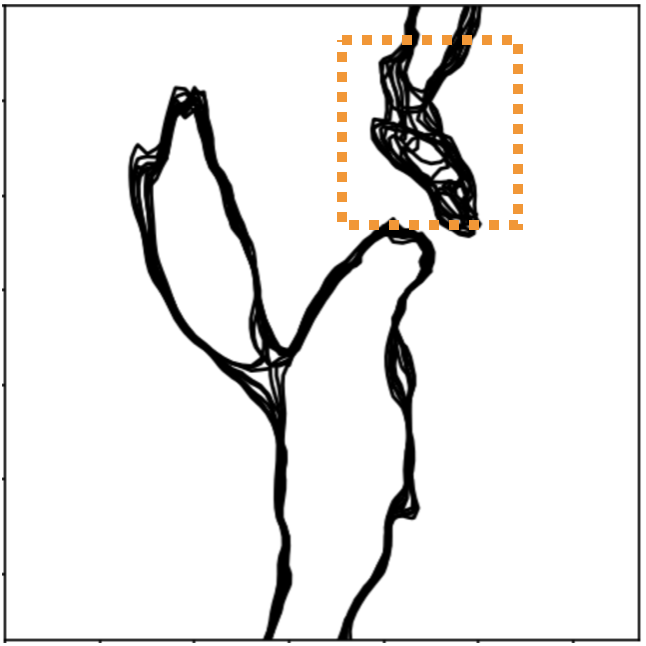}
\vspace{-2mm}
\caption{Spaghetti plot~\cite{potter2009}}
\label{fig:wind_spaghetti}
\end{subfigure}
\hspace{-2mm}
\begin{subfigure}{0.21\linewidth}
\centering
\includegraphics[width=0.8\linewidth]{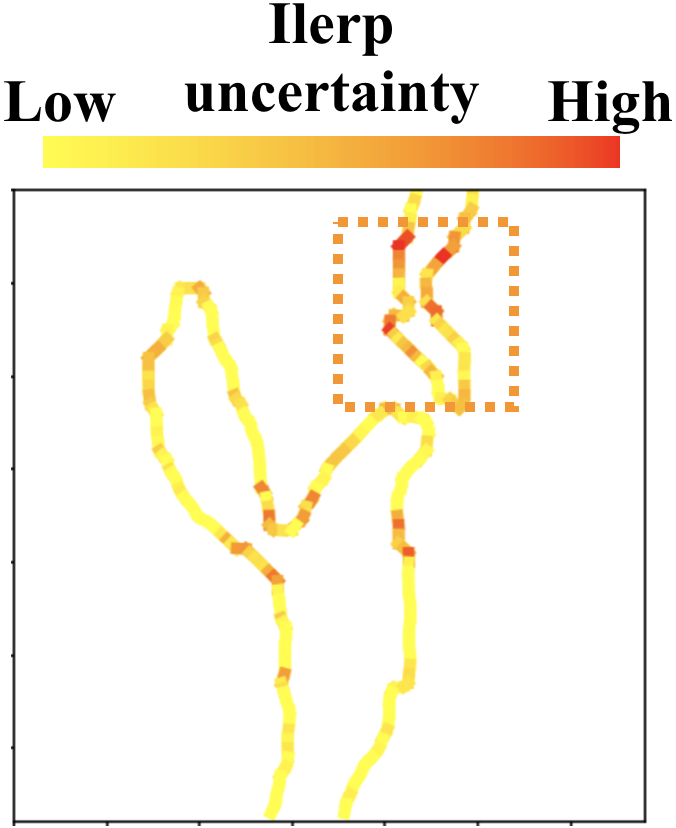}
\vspace{-2mm}
\caption{Ilerp uncertainty \\visualization~\cite{AthawaleEntezari2013}}
\label{fig:wind_ilerp}
\end{subfigure}
\hspace{-5mm}
\begin{subfigure}{0.22\linewidth}
\centering
\includegraphics[width=0.73\linewidth]{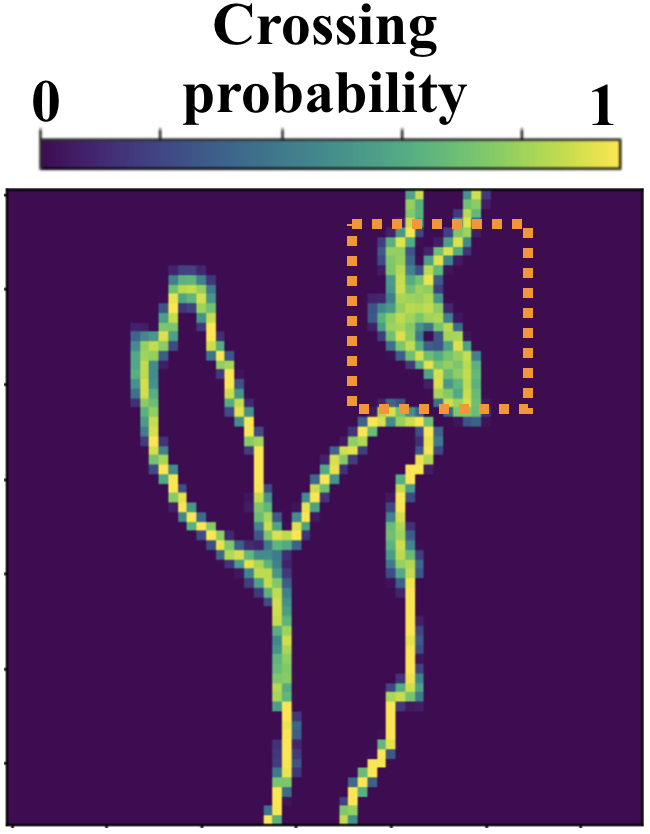}
\vspace{-2mm}
\caption{Probabilistic marching \\squares~\cite{PothkowWeberHege2011}}
\label{fig:wind_pms}
\end{subfigure}
\hspace{-10mm}
\begin{subfigure}{0.25\linewidth}
\centering
\includegraphics[width=0.66\linewidth]{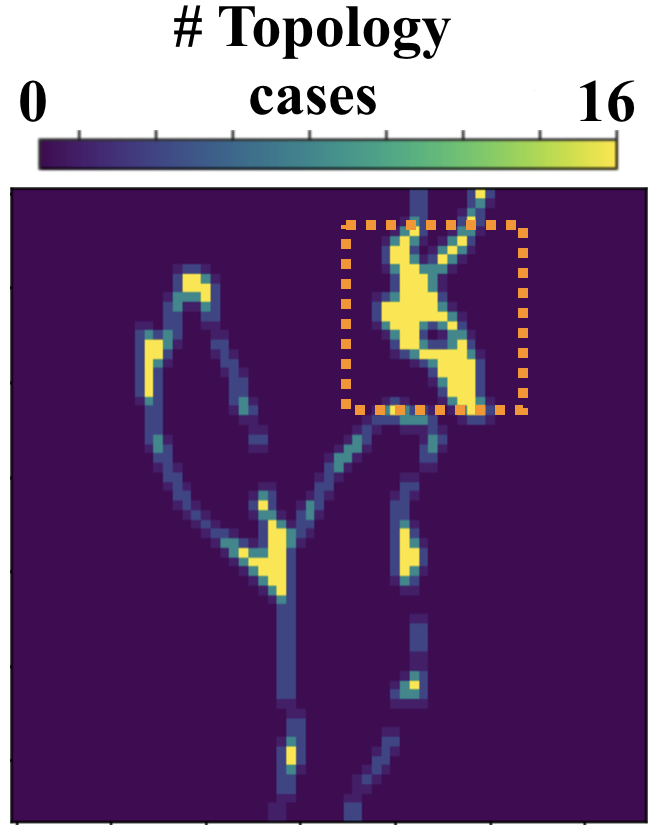}
\vspace{-2mm}
\caption{Topology case count \\(independent noise)}
\label{fig:wind_topoCountIndependent}
\end{subfigure}
\hspace{-10mm}
\begin{subfigure}{0.25\linewidth}
\centering
\includegraphics[width=0.65\linewidth]{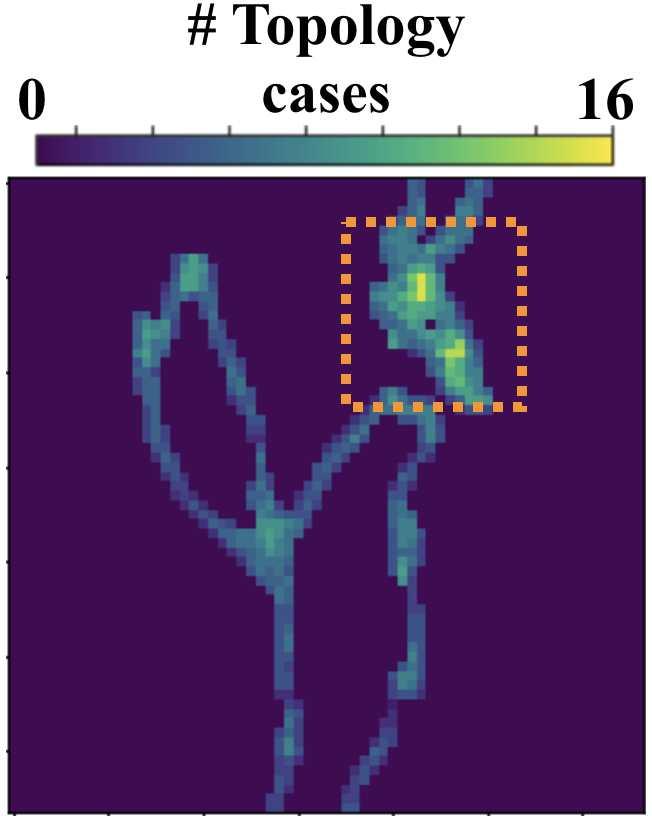}
\vspace{-2mm}
\caption{Topology case count \\(multivariate noise)}
\label{fig:wind_topoCountmultivariate}
\end{subfigure}

\vspace{2mm}

\hspace{-13mm}
\begin{subfigure}{0.29\linewidth}
\centering
\includegraphics[width=0.57\linewidth]{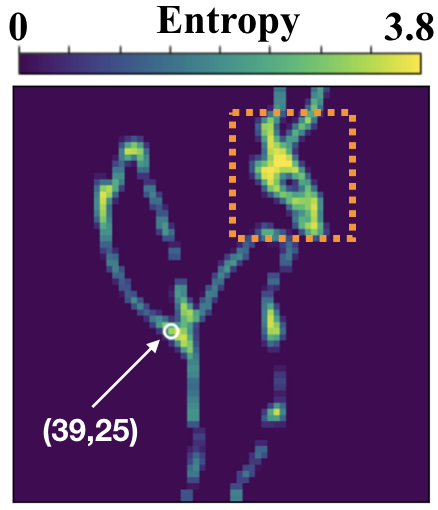}
\vspace{-2mm}
\caption{Entropy-based visualization\\ (independent noise)}
\label{wind_entropyIndependent}
\end{subfigure}
\hspace{-10mm}
\begin{subfigure}{0.3\linewidth}
\centering
\vspace{1mm}
\includegraphics[width=\linewidth]{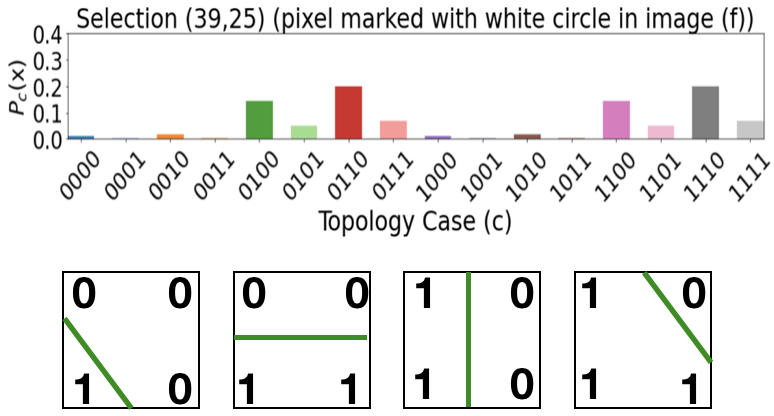}
\vspace{-2mm}
\caption{\fix{Topology distribution for\\ pixel (39,25) of image (f)}}
\label{wind_entropyIndependent}
\end{subfigure}
\hspace{-10mm}
\begin{subfigure}{0.29\linewidth}
\centering
\includegraphics[width=0.57\linewidth]{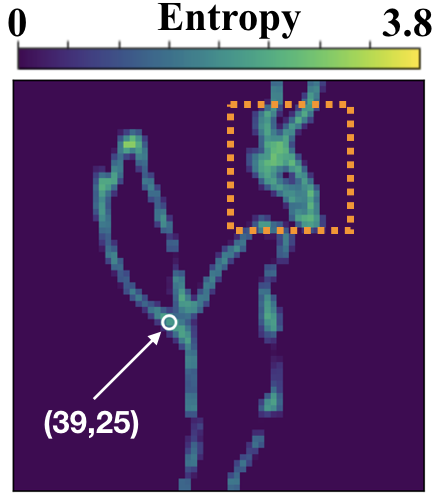}
\vspace{-2mm}
\caption{Entropy-based visualization\\ (multivariate noise)}
\label{wind_entropyMultivariate}
\end{subfigure}
\hspace{-10mm}
\begin{subfigure}{0.3\linewidth}
\centering
\includegraphics[width=\linewidth]{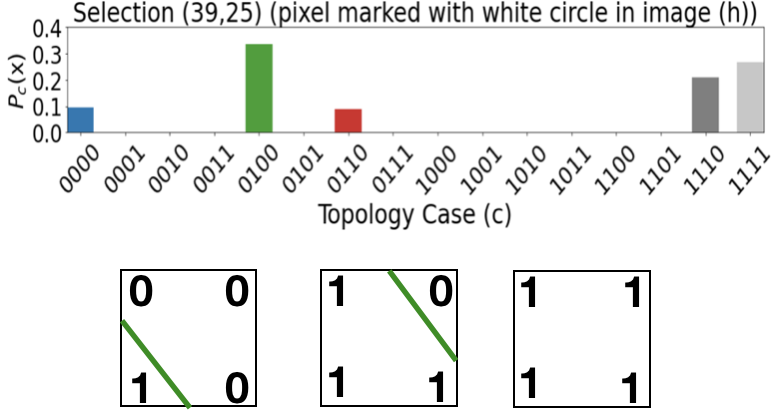}
\vspace{-2mm}
\caption{\fix{Topology distribution for\\ pixel (39,25) of image (h)}}
\label{wind_entropyMultivariate}
\end{subfigure}
\vspace{-2mm}
\caption{Uncertainty visualization of the wind dataset~\cite{Vitart2017} at $k=-40$ with noise modeling using independent Gaussians in images (b-d, f) and multivariate Gaussians in images (e,h). The reduced topology counts and entropy in images (e) and (h) compared to those in images (d) and (f), respectively, imply the more deterministic nature of the topology for multivariate noise models than independent noise models. In images (g,i), the high probability topology cases are depicted in the bottom rows. The vertex signs for any topology case (green segments) in images (g,i) are read from the top left corner of square cells in a counter-clockwise direction, in which 1 denotes a positive and 0 denotes a negative vertex.}
\label{fig:windVis}
\end{figure*}


\begin{figure*}[!ht]
\centering
\begin{subfigure}{0.33\linewidth}
\centering
\includegraphics[width=0.95\linewidth]{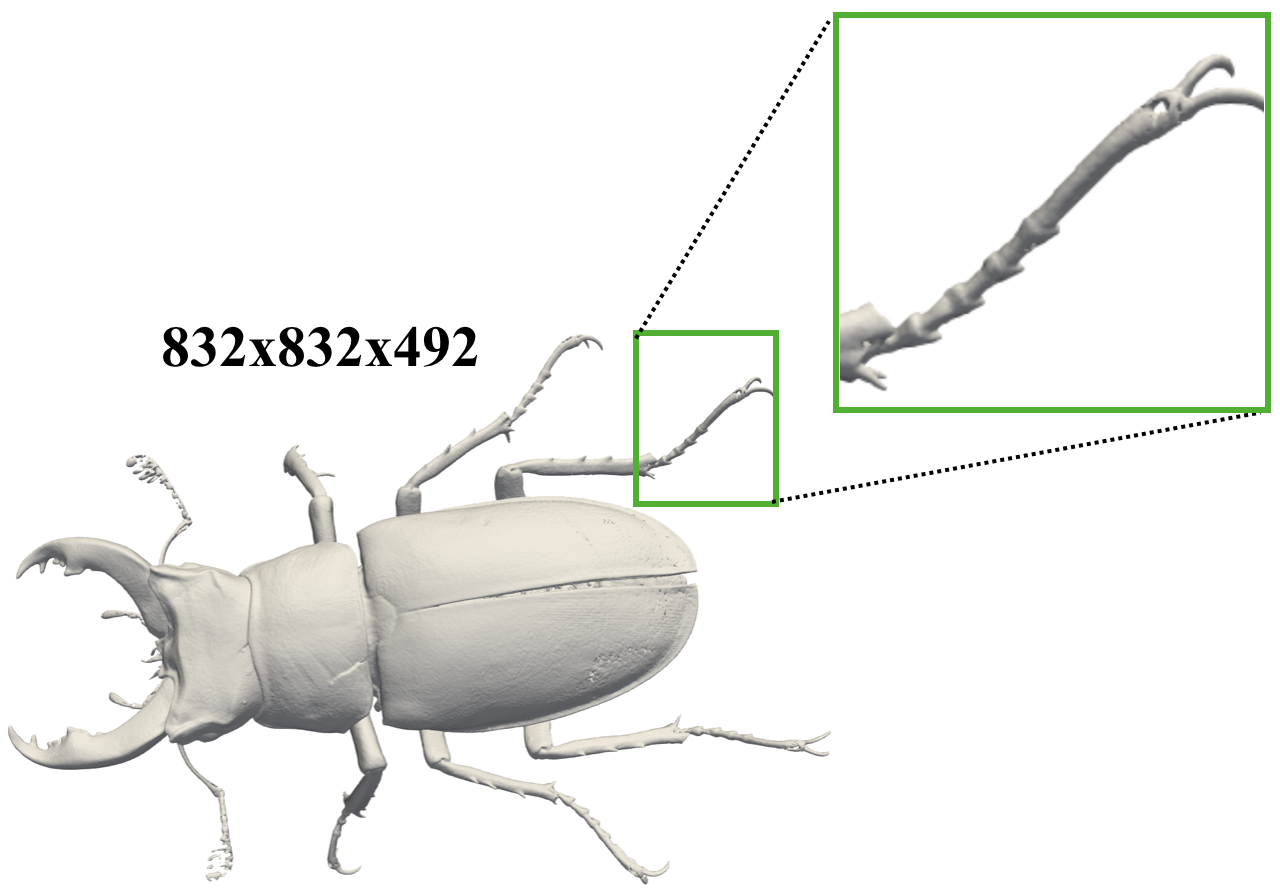}
\vspace{-2mm}
\caption{\fix{Level-set in a high-resolution image}}
\label{fig:beetle_gt}
\end{subfigure}
\begin{subfigure}{0.33\linewidth}
\centering
\includegraphics[width=0.95\linewidth]{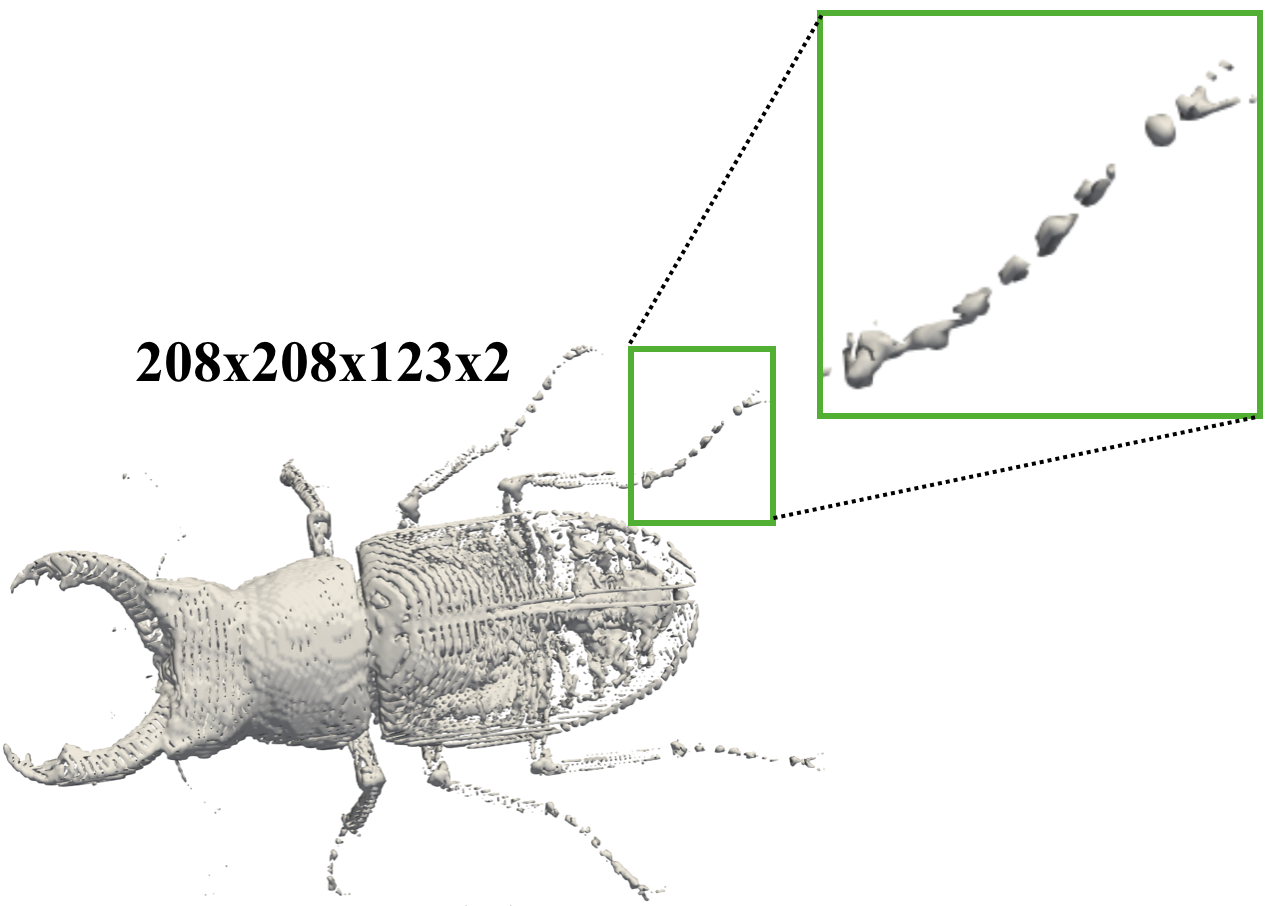}
\vspace{-2mm}
\caption{Most probable level-set topology~\cite{AthawaleEntezari2013}}
\label{fig:beetle_mostProbable}
\end{subfigure}
\begin{subfigure}{0.33\linewidth}
\centering
\includegraphics[width=0.95\linewidth]{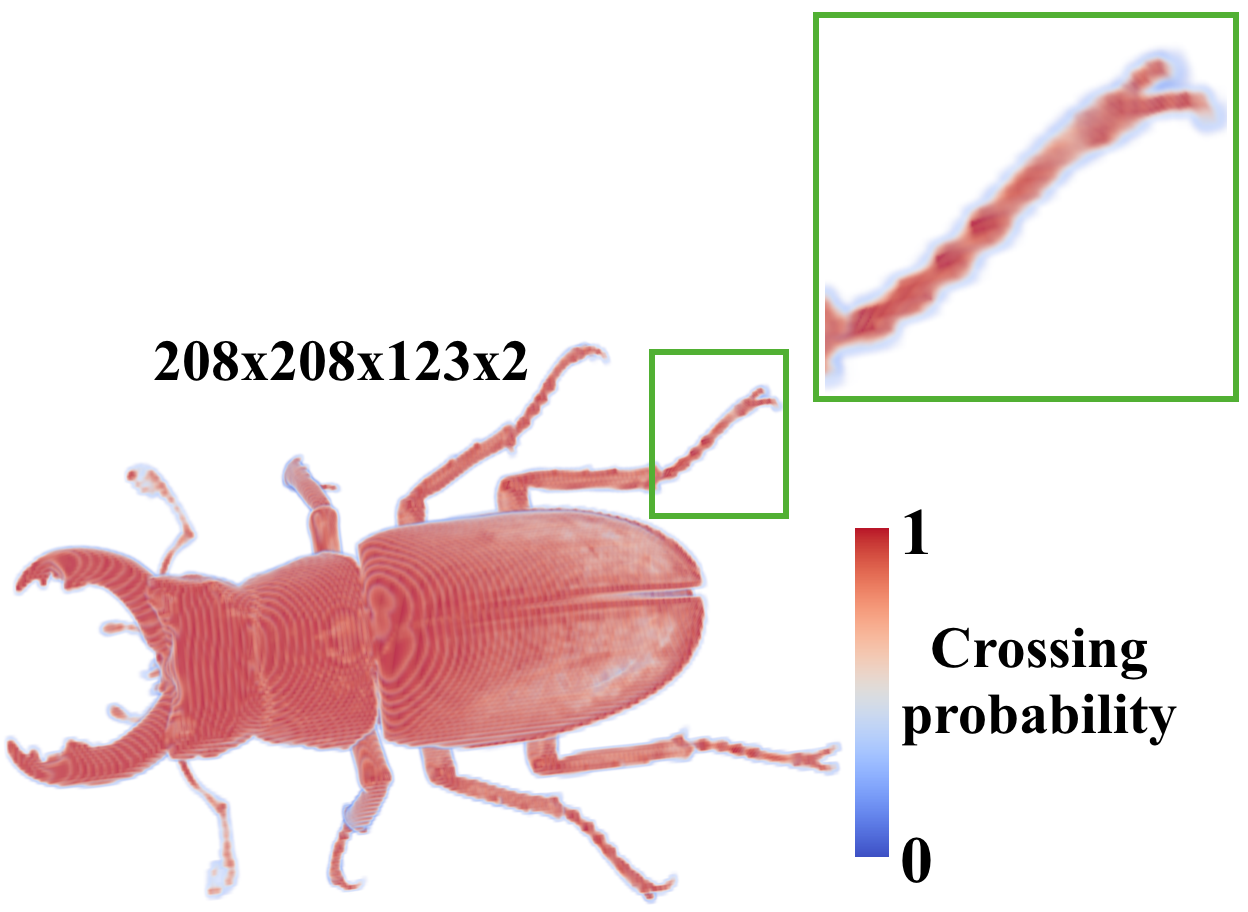}
\vspace{-2mm}
\caption{Probabilistic marching cubes~\cite{PothkowWeberHege2011}}
\label{fig:beetle_pmc}
\end{subfigure}

\vspace{2mm}

\begin{subfigure}{0.33\linewidth}
\centering
\includegraphics[width=0.95\linewidth]{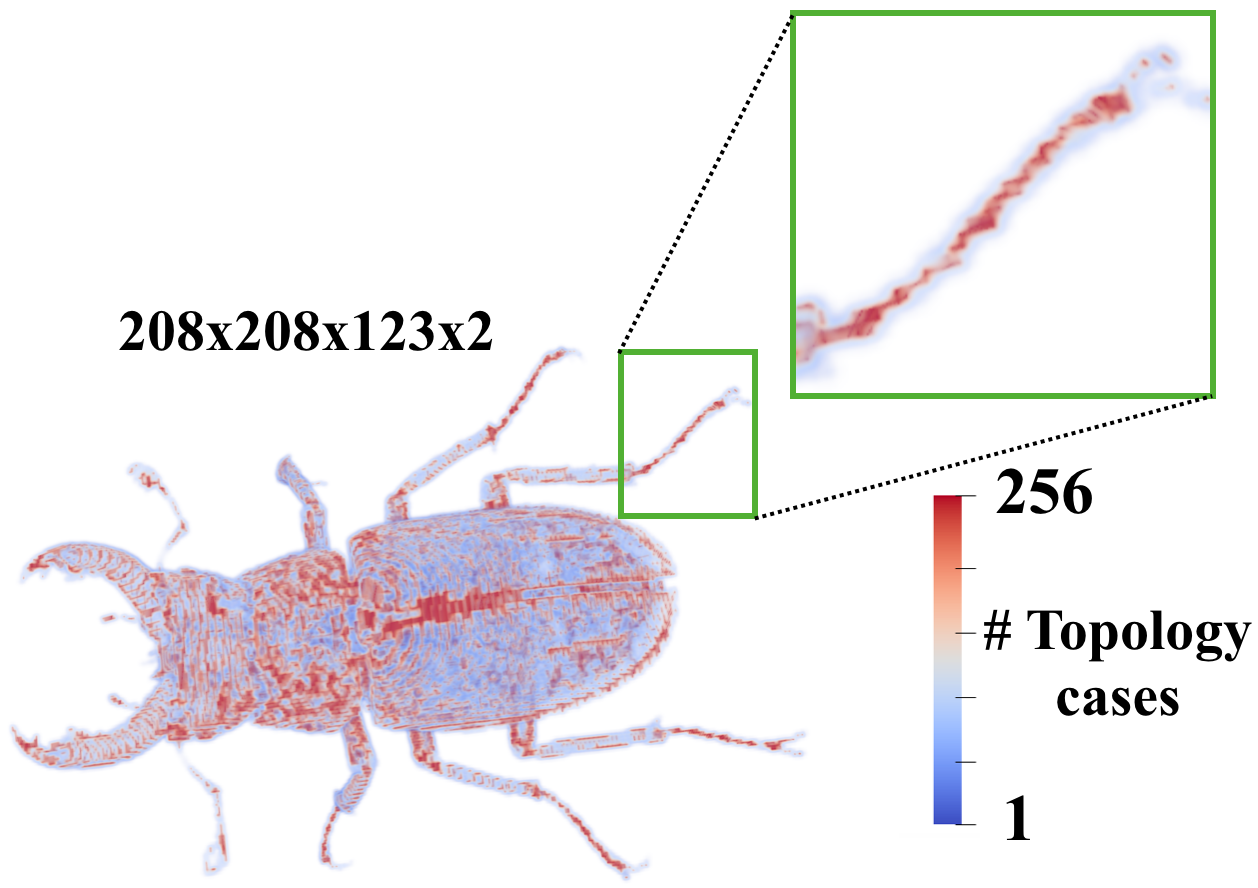}
\vspace{-2mm}
\caption{Topology case count visualization}
\label{fig:beetle_topoCount}
\end{subfigure}
\begin{subfigure}{0.33\linewidth}
\centering
\includegraphics[width=0.95\linewidth]{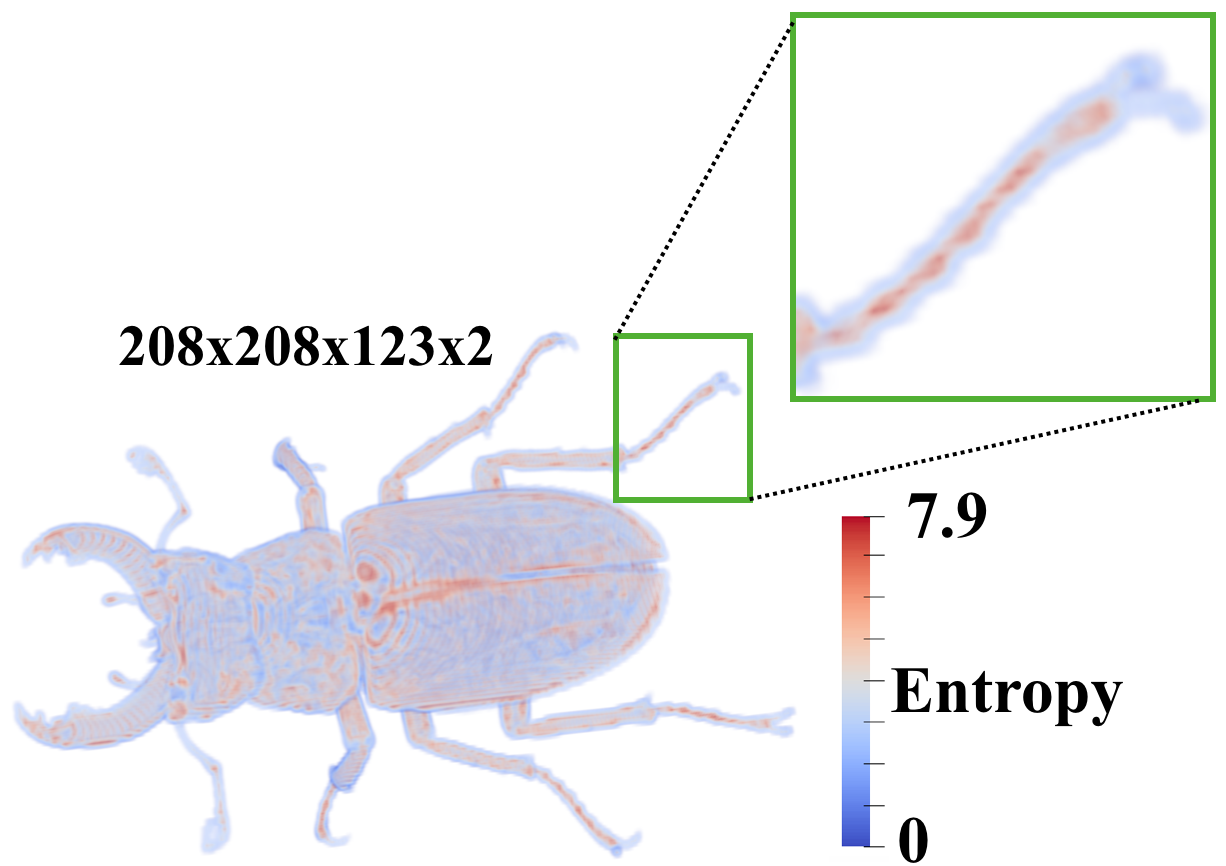}
\vspace{-2mm}
\caption{Entropy-based visualization}
\label{fig:beetle_entropy}
\end{subfigure}
\begin{subfigure}{0.33\linewidth}
\centering
\includegraphics[width=0.95\linewidth]{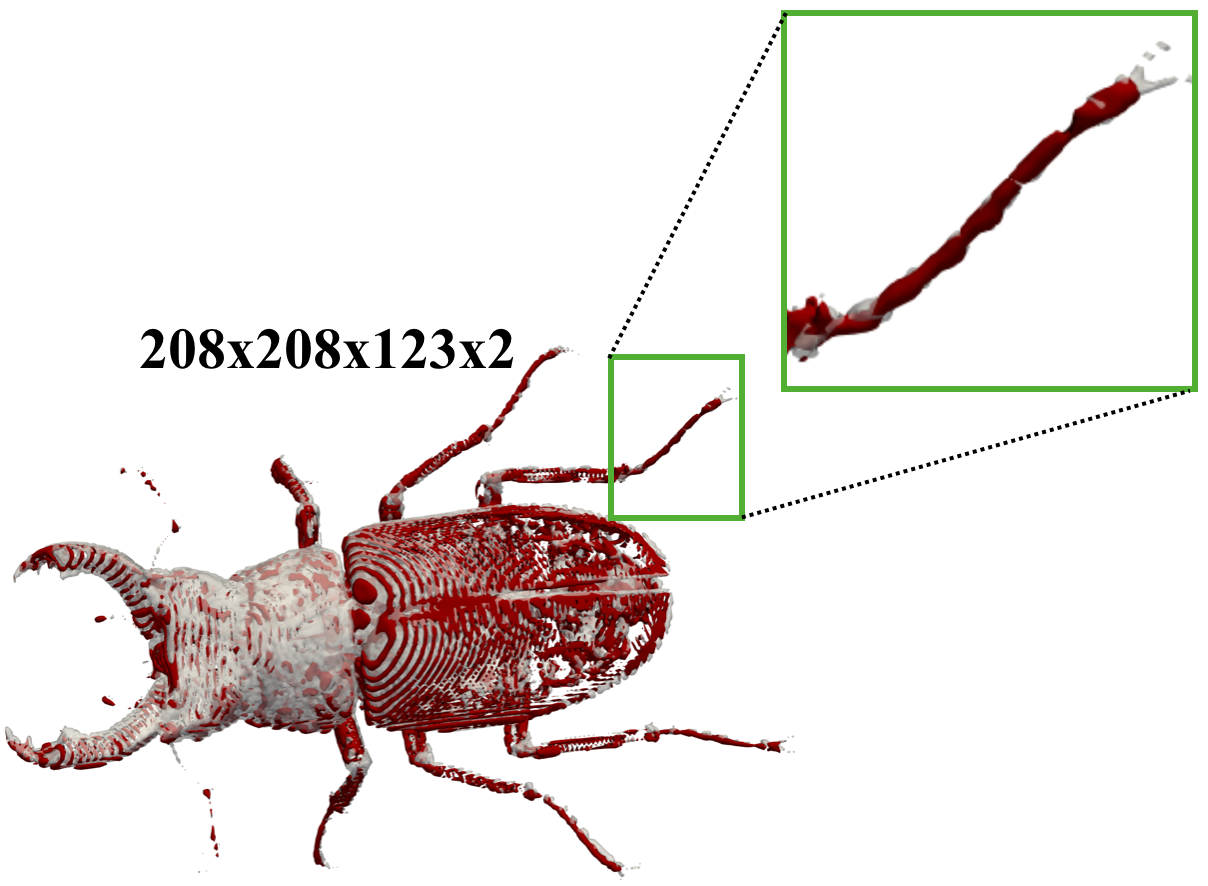}
\vspace{-2mm}
\caption{Entropy level-set (red)}
\label{fig:beetle_entropy}
\end{subfigure}
\vspace{-2mm}
\caption{Uncertainty visualizations for the stag beetle~\cite{ThompsonLevineBennett2011} hixel dataset at $k=900$. The noise in the data results in the breaking of the beetle leg in image (b). In probabilistic marching cubes, it is difficult to distinguish between the regions of high and topological uncertainty, which is easier using our visualizations in images (d-f). The relatively high sensitivity of the beetle leg topology to noise is detected in images (d-f) by the red regions. In image (f), the most probable level-set (gray) is overlaid with the entropy volume level-set (red) for entropy isovalue $5$.}
\label{fig:beetleVis}
\vspace{-4mm}
\end{figure*}


Fig.~\ref{fig:ackleyVis}b visualizes a spaghetti plot~\cite{potter2009} of level-sets, in which the orange and pink boxes enclose the positions of the relatively high and low spatial variability of the level-sets, respectively.
Fig.~\ref{fig:ackleyVis}c visualizes the most probable level-set extracted using the vertex-based classification (Sec.~\ref{sec:vbc}) with colormapping based on the Ilerp uncertainty~\cite{AthawaleEntezari2013}. The relatively high Ilerp variance (mapped to red) is oberved inside the orange box in Fig.~\ref{fig:ackleyVis}c. Fig.~\ref{fig:ackleyVis}d visualizes a result for the probabilistic marching squares~\cite{PothkowWeberHege2011} (Sec.~\ref{sec:pmc}), in which the level-crossing probabilities are colormapped.




Figs.~\ref{fig:ackleyVis}e-f visualize the results of our topology case count and entropy-based techniques, respectively (Sec.~\ref{sec:methods}). The yellow regions in Fig.~\ref{fig:ackleyVis}e indicate the cells that have $16$ possible MS topology cases with nonzero probability of occurrence, i.e., high topological uncertainty, across the ensemble. In Fig.~\ref{fig:ackleyVis}f, the high entropy mapped to yellow implies a relatively high level of randomness of level-set topology. Our proposed uncertainty visualizations clearly highlight the positions of relatively high topological uncertainty (yellow regions), which are not easily observed by visualizing the level-crossing probability (Fig.~\ref{fig:ackleyVis}d). Fig.~\ref{fig:entropyVsIsovalueAckleyVis} visualizes \fix{entropy boxplots for the isovalues sampled in the range $[-4.5, -2.5]$ for the Ackley ensemble. In Fig.~\ref{fig:entropyVsIsovalueAckleyVis}, the isovalues near $-4.5$ exhibit relatively higher median entropy (orange segments) than the isovalues near $-2.9$.}



In Fig.~\ref{fig:windVis}, we demonstrate the comparison of independent and correlated noise models and an application of interactive probability distribution queries~\cite{PotterKirbyXiu2012}. Specifically, we analyze the uncertainty in level-sets for the velocity magnitude fields derived from the wind ensemble dataset~\cite{Vitart2017} with $15$ ensemble members. Figs.~\ref{fig:windVis}a-c visualize the spaghetti plot, Ilerp uncertainty, and level-crossing probabilities similar to the visualizations in Figs.~\ref{fig:ackleyVis}b-d for the Ackley ensemble. 



Figs.~\ref{fig:windVis}d-e visualize the topology case counts for the independent and multivariate Gaussian noise models, respectively. The positions of relatively high topological variations are clearly highlighted (yellow regions) in Fig.~\ref{fig:windVis}d, which are not easily observed in Fig.~\ref{fig:windVis}c. The number of possible topological cases per cell with nonzero probability of occurrence is reduced or the topology becomes more deterministic for the multivariate Gaussian assumption in Fig.~\ref{fig:windVis}e. \fix{For the multivariate Gaussian assumption, we used a sample count $N=500$ for Monte Carlo sampling since increasing a sample count beyond $500$ did not visually alter the results significantly.} 

Fig.~\ref{fig:windVis}f and Fig.~\ref{fig:windVis}h visualize the entropy of the topological distributions for the independent and multivariate Gaussian noise assumptions, respectively. \fix{We investigate the probability distributions of the MS topology cases, interactively~\cite{PotterKirbyXiu2012}, at the pixels marked with white circles in Fig.~\ref{fig:windVis}f and Fig.~\ref{fig:windVis}h and visualize them in Fig.~\ref{fig:windVis}g and Fig.~\ref{fig:windVis}i, respectively.} In Fig.~\ref{fig:windVis}g, the high entropy of distributions is evidenced by the relatively high probability of four topology configurations $0100$, $0110$, $1100$, and $1110$, where $1$ denotes a positive and $0$ denotes a negative vertex. In contrast, the topology becomes more deterministic  in Fig.~\ref{fig:windVis}i with relatively high probability for the topology cases $0100$, $1110$, and $1111$. 




In Fig.~\ref{fig:beetleVis}, we apply our proposed uncertainty visualizations to a 3D hixel data~\cite{ThompsonLevineBennett2011} \fix{assuming the independent Gaussian-distributed uncertainty. The hixel technique produces a reduced representation of the original data by partitioning these data into blocks and summarizing each block with a probability distribution.} In our example, we derived the hixel data from the stag beetle dataset~\cite{dataset-stagbeetle} with resolution $832 \times 832 \times 492$. For the hixel-based representation, we partitioned the dataset into blocks of size $4 \times 4 \times 4$ and summarized each block with a Gaussian distribution. The hixel dataset, therefore, has a resolution of $208 \times 208 \times 123 \times 2$, where each block stores the mean and standard deviation of a Gaussian distribution.

The hixel-based reduced representation comes at the cost of increased uncertainty in the data and, hence, the level-set positions. Fig.~\ref{fig:beetleVis}a visualizes the level-set extracted from the original high-resolution stag beetle dataset at isovalue $k=900$. Fig.~\ref{fig:beetleVis}b visualizes the most probable isosurface extracted using the vertex-based classification~\cite{AthawaleEntezari2013} for the hixel dataset. The relatively high sensitivity of the beetle leg topology to noise results in the breaking of the beetle leg in Fig.~\ref{fig:beetleVis}b. Fig.~\ref{fig:beetleVis}c visualizes the result for the probabilistic marching cubes, in which the positions of high and low topological uncertainty are not clearly separated. Our proposed uncertainty visualizations in Fig.~\ref{fig:beetleVis}d-f clearly detect the relatively high sensitivity of the beetle leg topology to noise, as visualized with the red regions. In Fig.~\ref{fig:beetleVis}f, we overlay the most probable level-set (gray) with the level-set extracted from the entropy volume (red) for the entropy isovalue $5$. Thus, the red isosurface encloses the positions that have relatively high topological uncertainty.  




\section{Conclusion and Future Work}
\label{sec:conclusion} 
In this paper, we study uncertainty arising in the topology cases of the MS and MC algorithms for level-set visualizations when the data uncertainty is modeled with independent and correlated noise distributions. Specifically, we propose the topology case count and entropy-based techniques for uncertainty quantification and visualization of the topology cases. We demonstrate the effectiveness of our proposed uncertainty visualizations by comparing the results with previously proposed spaghetti plots~\cite{potter2009}, probabilistic marching cubes~\cite{PothkowWeberHege2011}, and Ilerp uncertainty visualizations~\cite{AthawaleEntezari2013}. 

\fix{Our proposed uncertainty quantification framework has a few limitations. First, we assume no correlation among the 16 MS (or 256 MC) topology cases that are distinguished based on the cell vertex signs. A few topology cases, however, are rotated or flipped versions of other MS topology cases. Considering such correlations for uncertainty quantification could be interesting future work. Further, a specific combination of cell vertex signs may correspond to multiple possible topologies within a grid cell~\cite{Nielson:1991:TAD, Nielson:2003:OMC}, which we plan to take into account for uncertainty quantification in the future.} Currently, we limit interactive probability queries~\cite{PotterKirbyXiu2012} to explore the MS topological uncertainty. Applying such a framework to the MC topology cases, however, is nontrivial and impractical. We would like to study the MC topology cases further. In our study, we restrict the topology uncertainty analysis for each cell of a scalar grid similar to the MS and MC algorithms. We would like to expand this analysis to take into account correlations with a local neighborhood. 


\acknowledgments{
This work was supported in part by
a grant from the Intel Graphics and Visualization Institutes of XeLLENCE. \fix{We would like to thank the reviewers of the paper for their valuable feedback.}} 

\bibliographystyle{abbrv-doi}

\bibliography{mcTopologyUncertaintyVis}
\end{document}